\documentclass[twocolumn,showpacs,preprintnumbers,amsmath,amssymb]{revtex4}

\usepackage{graphicx}

\begin{document}
\title{In-situ Observation of Incompressible Mott-Insulating Domains \\of Ultracold Atomic Gases}
\author{Nathan Gemelke}
\author{Xibo Zhang}
\author{Chen-Lung Hung}
\author{Cheng Chin}

\address{The James Franck Institute and Department of Physics, \\ University of Chicago, Chicago, IL 60637, USA}

\date{April 9, 2009}

\begin{abstract}We present a direct measurement of the density profile of a two-dimensional Mott Insulator formed by ultracold atoms in an optical lattice. High resolution absorption imaging is used to probe the ``wedding-cake'' structure of a trapped gas as it crosses the boundary from a unit-filled Mott insulating phase to the superfluid phase at finite temperature.  Detailed analysis of images yields measurements of temperature and local compressibility; for the latter we observe a strong suppression deep in the Mott-insulating phase, which is recovered for the superfluid and normal phases.  Furthermore, we measure spatially resolved fluctuations in the local density, showing a suppression of fluctuations in the insulator. Results are consistent with the fluctuation-dissipation theorem for insulator, superfluid and normal gas.
\end{abstract}
\pacs{73.43.Nq,03.75.Kk,05.30.Jp,03.75.Lm}

\maketitle

The observation of the Superfluid (SF) to Mott-insulator (MI)  phase
transition of ultracold atoms in optical lattices \cite{bloch_mottins} marked an experimental
breakthrough in the study of many-body physics in an atomic system. This
development provided the first tangible example of a quantum phase
transition (one that occurs even at zero temperature) and suggested a
highly correlated and gapped Mott insulating state of a Bose gas, which
persists down to zero temperature without Bose-condensation
and macroscopic phase coherence. Since its theoretical inception \cite{kaganov87,fisher89,Jaksch98}, two of the most celebrated properties of the bosonic Mott insulator have been its incompressibility and suppression of
local density fluctuations \cite{svistunov07}, induced by enhanced inter-particle
interaction and reduced mobility in optical lattices. The result
for a trapped atom gas, where the local chemical potential varies in
space, is the remarkable ``wedding-cake'' density
profile, where successive MI domains are manifest as plateaus of
constant density.  Related phenomena have
been studied through the coherence \cite{bloch_mottins,spielman08},
transport \cite{bloch_mottins,Kohl05}, noise correlations \cite{Folling05}, and number variance
\cite{greiner02,gerbier06}, but direct observation of the
incompressibility has proven difficult due to the inhomogeneous
nature of all experiments to date, and to the technical difficulty
of spatially resolved measurements. Innovative experimental
efforts have yielded evidence \cite{folling06,Campbell06}
that these plateaus exist, though none has directly
observed this effect by imaging a single physical system in situ.

We report studies based on direct in-situ imaging of an atomic MI. By loading a degenerate Bose gas of
cesium-133 atoms into a thin layer of a two-dimensional optical
lattice potential, and adiabatically increasing the optical lattice
depth, we observe the emergence of an extremely flat density
near the center of the cloud, which corresponds to a
MI phase with accurately one atom-per-site. From density
profiles, we extract important thermodynamic and statistical
information, confirming the incompressibility and reduction of
density fluctuations in the MI as described by the
fluctuation-dissipation theorem.

The single layer, two-dimensional (2D) optical lattice is
formed by two pairs of counter-propagating laser beams derived from
a Yb fiber laser at wavelength $\lambda=1064$~nm. The pairs are
oriented orthogonally on the horizontal ($x-y$) plane,
forming a square lattice with site spacing $d=\lambda/2=0.532$~$\mu$m. A weak harmonic potential of
$V_H=m(\omega_x^2x^2+\omega_y^2y^2)/2$ localizes the sample,
where $m$ is the cesium mass, and the geometric mean of the trap
frequencies is $\omega_r=\sqrt{\omega_x\omega_y}=2\pi\times
9.5(1+V/82E_R)$Hz; we have included a weak dependence on
the lattice depth $V$ ($E_R=h\ 1.3$kHz is the lattice recoil energy
and $h$ is Planck's constant.) Vertical confinement is
provided by an additional vertical optical lattice with a site spacing
$4~\mu$m, formed by two beams intersecting at an
angle of $15^\circ$, confining atoms to an oscillator
length $a_z=0.30$~$\mu$m. The sample is loaded into a single
site of the vertical lattice, kept deep to prevent vertical tunneling. Tunneling in the horizontal
2D lattice is controlled by varying the lattice depth $V$ \cite{bloch_mottins}.  Details on preparation of the atomic sample can be found in the supporting material and Ref. \cite{Hung08}.

We obtain a top view of the sample using absorption imaging, directly revealing the atomic surface density $n(x,y)$ on the horizontal plane. The imaging resolution is $3\sim4\mu$m, and magnification such that one imaging pixel corresponds to an area of ($2\mu$m$)^2$ on
the object plane. Conveniently, unit filling in a 2D optical lattice has a
moderate optical density of $OD=O(1)$ on resonance.

\begin{figure}[t]\center
\includegraphics[width=3.25 in]{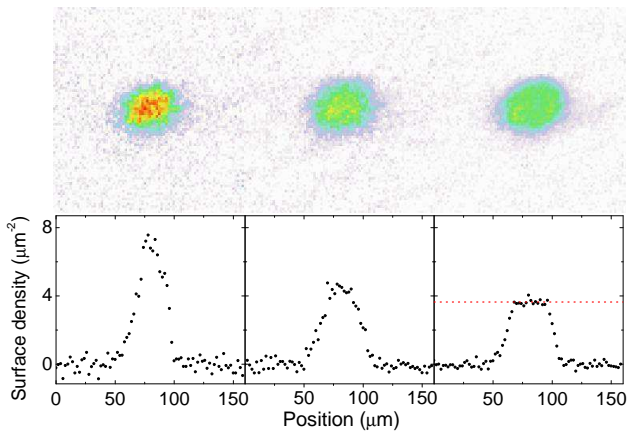}
\caption{False color density profiles and line cuts of $N=7500$
ultracold cesium atoms at scattering length $a=310$~$a_B$ in 2D
optical lattices. (A) Superfluid regime (shallow lattice
$V=0.14E_R$), (B) Phase transition regime (medium lattice depth
$V=7.2E_R$), and (C) Mott insulator regime (deep lattice
$V=38E_R$)  The red dashed line indicates the expected density for one atom-per-site. Line cuts pass through the center of the sample. } \label{fig1}
\end{figure}

The superflulid-to-Mott insulator (SF-MI) transition of ultracold
atoms in an optical lattice is described by the Bose-Hubbard model,
characterized by on-site interaction $U$ and the tunneling $J$ \cite{Jaksch98}. In
2D optical lattices, superfluid is converted into a MI
when $U/J$ exceeds 16 \cite{spielman07,spielman08}. Here, the
SF-MI phase transition can be induced by either increasing the
lattice potential $V$ \cite{bloch_mottins,spielman08,Campbell06} or the atomic scattering length $a$
via a magnetically-tuned Feshbach resonance \cite{chin08}, together
providing complete, independent control of $U$ and $J$.

Atomic density profiles in the lattice are shown in
Fig.~\ref{fig1}. For weak lattice depths (superfluid regime), the
density profiles are bell-shape, with negative curvature at the
center (Fig.~\ref{fig1}A), indicating a
finite, positive compressibility dictated by the interaction
coupling constant (discussed below.) In sufficiently deep
lattices, we observe a flattened density at the
center of the sample (Fig. \ref{fig1}B, C), indicating development of a Mott insulating phase with
one particle per lattice site. This density plateau, an important feature of the MI phase, arises from incompressibility.

A primary check on the MI is to
compare the measured density in the plateau to that corresponding to one atom-per-site, given by MI physics as a ``standard candle'' of atomic density.  Using the known scattering cross-section, correcting for saturation effects (see Methods), we determine the plateau density to be $n=3.5(3)/\mu$m$^2$, in agreement with the expected value $1/d^2=3.53/\mu$m$^2$.

\begin{figure}[t]%\center
\includegraphics[width=3 in]{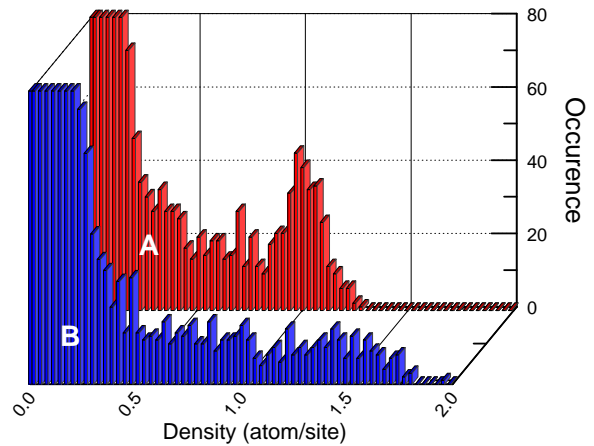}
\caption{Histograms of density profiles in the MI regime
(A, $V=38E_R$, $a=460a_B$) and the superfluid regime (B,
$V=0.5E_R$, $a=460a_B$.) The histograms are based on an average of
three density images. The bin size is $\Delta n=0.03$.} \label{fig2}
\end{figure}

To distinguish a MI from superfluid or
normal gas, we histogram the occurrence of pixels $h(n)$ in the images corresponding
to a density $n$ with a bin size of
$\Delta n\ll n$. The MI plateau,
containing a large number of pixels with similar atomic density,
appears as a peak at $n=1/d^2$ (Fig.~\ref{fig2}A). In general, the occurrence
of a particular density $n$ can be regarded as the rate at which
local chemical potential changes with density, multiplied by the
number of pixels $w(\mu)\Delta\mu$ corresponding to a chemical
potential between $\mu$ and $\mu+\Delta\mu$. The occurrence at density $n$ is then $h(n)=\Delta
n\,w(\mu)\Delta\mu/\Delta n\approx \Delta n\,w(\mu)\kappa^{-1}$,
where $\kappa=\partial n/\partial \mu$ is the local compressibility \cite{Rigol08}. In a harmonic trap, $w(\mu)=2\pi/md^2\omega_r^2$ is constant, and the histogram is a particularly useful tool to
distinguish different phases. For a pure BEC in the Thomas-Fermi
limit, the compressibility is constant to the
maximally-allowed density $n_{pk}$, and results in a constant $h(n)$
for $n\le n_{pk}$ (see Fig.~\ref{fig2}B for
$0.5/d^{2}<n<1.5/d^{2}$). For the MI, the density is
insensitive to chemical potential in a narrow range near $n=1/d^{2}$, indicating a vanishing compressibility,
and thus a sharp histogram peak at $n=1/d^{2}$. The peak's presence in
Fig.~\ref{fig2}A is thus directly related to the incompressibility
in the Mott phase. Finally, the
compressibility of a normal (ideal) gas is proportional to its
density, thus $h(n)\propto 1/n$, leading to the strong upturn
at low densities in Fig. \ref{fig2}A,B for both regimes.

\begin{figure}[t]%\center
\includegraphics[width=3 in]{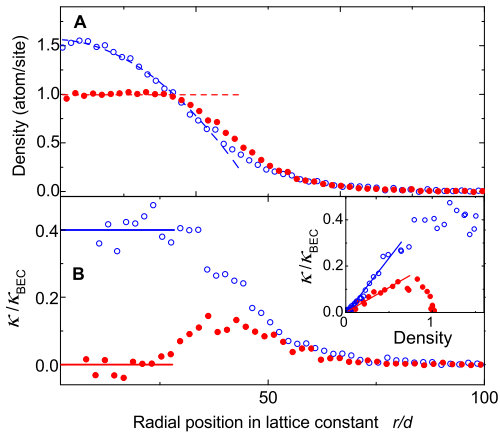}
\caption{(A) Radially averaged profiles (3 images) in the superfluid
(blue open circles: $V=0.25E_R$) and MI (red circles:
$V=38E_R$) regimes. Particle number and scattering length are
$N=8100$ and $a=460a_B$, respectively. A quadratic fit to the center
portion of the sample extracts the curvature near $r=0$. (B)
Normalized compressibilities derived from (A) using Eq.~(1) in the superfluid
(blue open circles) and MI (red circles) regimes. The
horizontal lines indicate the compressibility near $r=0$, estimated
from the quadratic fits in (A). The rising compressibility
at $r=40d$ marks the MI boundary. The inset shows the
dependence of compressibility on atomic density. The linear
dependence at low densities is fit by the solid lines.} \label{fig3}
\end{figure}

Much more information can be obtained from the density profiles, as recently suggested in
Ref.~\cite{Ho_2008}. For example, the compressibility in a two-dimensional cylindrically symmetric trap can be written $\kappa=\partial n/\partial \mu=-n'(r)/(rm\omega_r^2)$, where we have assumed the local density
approximation, and that the chemical potential depends on the trapping
potential $\mu=\mu_0-V_{H}(r)$. For a BEC in the Thomas-Fermi
regime, the compressibility is a positive and constant
$\kappa_{BEC}=1/g$, where $g=\sqrt{8\pi} a\hbar^2/ma_z$ is the (2D)
interaction parameter \cite{Tanatar02}. We can thereby
relate the measured compressibility to that of a BEC as

\begin{equation}
\frac{\kappa}{\kappa_{BEC}}=-(\frac2{\pi})^{7/2}\frac{n'(r)}{rd^{-4}}\frac
a{a_z}(\frac{E_R}{\hbar\omega_r})^2. \label{kappa}
\end{equation}

\noindent

We evaluate $\kappa$ from azimuthally averaged density profiles
(Fig.~\ref{fig3} A). Eccentricity of the trap is corrected by
rescaling the principal axes. Due to the singular nature of
$n'(r)/r$ near the center, we evaluate $\kappa$ there by
fitting $n(r)$ to a quadratic, $n(r)=n(0)-\alpha r^2$. The
curvature then gives the compressibility as
$\kappa(0)=2\alpha/m\omega_r^2$, for which we obtain
$\kappa/\kappa_{BEC}=0.4(1)$ in a weak lattice and
$\kappa/\kappa_{BEC}=0.006(6)$ in a strong lattice (See
Fig.~\ref{fig3}). In the weak lattice (SF regime), the finite and
constant compressibility at the center agrees with expectation for
the superfluid phase, though lower than expected, which we attribute to
finite temperature and calibration of trap
parameters. The finite temperature is also clear in the exponential
tail of the density profile and the compressibility, from
which we derive the temperature $11(2)$~nK in the
superfluid regime and $24(4)$~nK in the MI regime.

In a deep lattice (MI regime), we observe a strong reduction of
the compressibility in the trap center, below that in the superfluid phase for the weak lattice, strongly supporting the emergence of a MI phase at the
center of the sample. Away from center, $\kappa$ suddenly
increases at $r=25d$, then decreases for $r>50d$. The exponential
decay is again consistent with a normal gas. Between MI and normal gas ($25d<r<50d$), a more detailed
measurement and model of compressibility would be necessary to
identify the local phase.

\begin{figure}[t]%\center
\includegraphics[width=3.5 in]{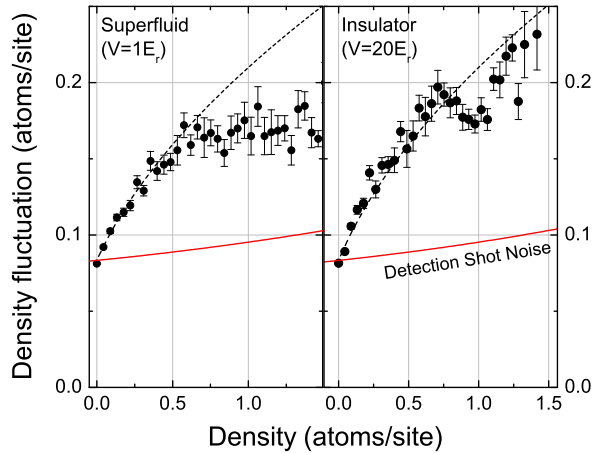}
\caption{The fluctuation of local density extracted from a set of
twelve absorption images in the deep (MI) and weak (SF) lattice regimes. The insulator and superfluid show a pronounced difference at the density of one atom-per-site, where the insulator's fluctuation is suppressed by its incompressibility.  In the superfluid phase, constant compressibility initiates a flattening of fluctuations with density.  At low densities, in both regimes, the fluctuation shows a characteristic $\sqrt{n}$ dependence. The dashed line shows best fit $\sqrt{n}$ dependence for the normal gas. The total number of atoms was $N=8300$ (SF) and $N=9600$ (MI) with $a=310a_B$ for both sets.} \label{fig4}
\end{figure}

Incompressibility necessarily implies a low
particle number fluctuation; this relationship, a result of the fluctuation-dissipation theorem (FDT) (see e.g. \cite{Batrouni02,Huang}), takes the form

\begin{equation}\label{eqn2}
\delta n ^2 = \kappa\,k_B T
\end{equation}

Resolved in-situ imaging provides an enticing opportunity to
measure fluctuations of the local density \cite{Esteve06,svistunov07}, and thus check the
validity of the FDT.  We measured fluctuations by recording multiple absorption images,
calculating the variance of density measured in each pixel (each collects signals from a patch of $(2\mu m/d)^2\approx14$ lattice sites). Fig.~\ref{fig4} shows
the recorded fluctuations, where pixels are binned
according to their mean density. Fluctuations consist of
detection (photo-electron shot) noise and thermal and quantum atomic density
fluctuations. Detection shot noise can be well-calibrated and
modeled by analyzing portions of the images with low density;
extension to higher optical depth (density) shows the weak
dependence illustrated (Fig. \ref{fig4}).

Above the detection noise, density
fluctuations (see Fig.~\ref{fig4}) show
a strong qualitative agreement with the compressibility presented
in Figure \ref{fig3} as expected from the FDT.  For example, the Mott-phase shows a strong
suppression of fluctuations at the density of one atom-per-site.
The superfluid regime lacks this feature, instead
showing a pronounced flattening as the sample transitions from normal gas
to superfluid, as expected from the constant compressibility in the
superfluid phase (Figure \ref{fig3}, inset).  Finally, at low density, the normal gas shows a temperature-independent fluctuation of
$\delta n=\gamma \sqrt{n}$, which can be anticipated from the
inset of Figure \ref{fig3}, and agrees with the FDT.  The coefficient $\gamma$ is roughly consistent with the FDT, and measured imaging resolution (see methods).

Clearly, in situ imaging of the Mott insulator is a powerful new tool to investigate
new quantum phases of cold atoms in optical lattices. From the
density profiles, not only can one observe the density plateau,
incompressibility and reduction of fluctuations in the Mott
insulating phase, but also demonstrate a qualitative validation of the fluctuation-dissipation theorem. Relatively modest extension of this work holds new promise for studying the role of quantum
fluctuations, correlation and thermodynamics near a quantum phase
transition.

We thank T.L. Ho, R. Scalettar, E. Mueller, and R. Hulet for helpful discussion.  This work is supported by NSF No. PHY-0747907, NSF-MRSEC DMR-0213745, and ARO No. W911NF0710576 with funds from the DARPA OLE Program. N.G. acknowledges support from the Grainger Foundation.

\bibliography{plateau}

\clearpage

\begin{center}
\section*{Supporting Material}
\end{center}

\subsection*{Preparation of BEC in a thin 2D optical lattice}

\begin{figure}[h]\center
\includegraphics[width=3.5 in]{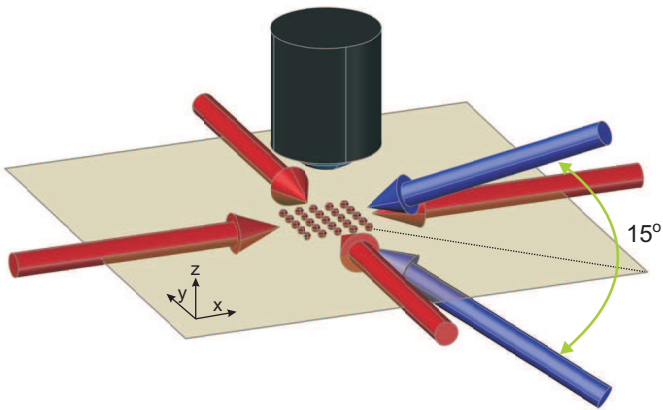}
\caption{Experimental apparatus used to investigate quantum phases in two-dimensional optical lattices.  Two pairs of counterpropagating beams (red) form a 2D square optical lattice, into which a cesium Bose-condensate is loaded.  Atoms are confined in the vertical direction by loading into a single site of a separate one-dimensional lattice in the vertical direction, formed by two beams (blue) intersecting at a small angle.  A high numerical aperture imaging lens images the shadow (absorption) cast by the atoms illuminated by imaging light propagating vertically.} \label{fig5}
\end{figure}

The $^{133}$Cs BEC is formed in a crossed-beam dipole trap by an
efficient evaporative cooling method [1]. The dipole trap consists
of three beams on the horizontal plane: two orthogonal beams at the
wavelength of 1064~nm (Yb fiber laser, YLR-20-1064-LP-SF, IPG),
focused to $1/e^2$ radii of 350~$\mu$m, and one CO$_2$ laser beam (not shown in Fig \ref{fig5}) at
the wavelength of 10.6~$\mu$m (Gem-Select 100, Coherent), focused to
a vertical $1/e^2$ radius of 70~$\mu$m and horizontal of 2~mm (see Fig. \ref{fig5}). The
CO$_2$ beam intersects the Yb laser beams at an angle of 45$^\circ$ and
provides an enhanced vertical confinement to support the atoms
against gravity. With $N=10^4$ atoms in a pure condensate, the
Thomas-Fermi radii of the condensate are $(r_x,r_y,r_z)=(23,
14, 3.6)\mu$m.

After a pure BEC is obtained, the sample is compressed
vertically by introducing a vertical lattice, formed by two laser
beams (Innolight Mephisto) inclined at +7.5$^\circ$ and -7.5$^\circ$ relative to
the horizontal plane. The vertical lattice has a spacing of 4~$\mu$m
and, together with the crossed dipole trap, forms an array of 2D oblate ``pancake''
potentials.

In order to load the condensate into a single pancake trap, we
first ramp the magnetic field to 17.2~G in 400~ms, reducing the
s-wave scattering length to $a<$10~$a_B$, and then turn on the
vertical lattice in 100~ms.  Atomic
population in other lattice sites, if any, can be identified by observing
an interference pattern in time-of-flight images taken from the side. For this work, we
observe a sufficiently weak interference pattern contrast to conclude
$>98\%$ of the atoms are in a single pancake trap. After the
vertical lattice is fully turned on, the CO$_2$ laser intensity is
ramped to zero in 100~ms and the scattering length ramped to a
desired value by tuning the external magnetic field.

The 2D lattice potential in the horizontal ($x$- and $y$-)
directions is formed by introducing retro-reflections of the 1064~nm
dipole trap beams. A continuous evolution from a pure dipole trap
(with zero retro-reflection) to a 2D optical lattice (with
significant retro-reflection) is achieved by passing each dipole
trap beam (after it passes through the atomic cloud once) through two acousto-optic modulators (AOMs)
controlled by the same radio-frequency (rf) source, then off a retroreflection mirror. The AOMs induce
an overall zero frequency shift, but permit a dynamic control of the
retroreflection intensity over six orders of magnitude. Onsite
interaction energy $U$ and tunneling rate $J$ are
evaluated from the measurements of the lattice vibration frequencies
and the band structure calculation.

\subsection*{Calibration of atomic surface density}

By varying the intensity of the imaging beam, we measure the optical
depth in the density plateau using $OD=\ln(M_0/M)$, where $M$ is the
number of photons collected by a CCD pixel in the presence of the atoms
and $M_0$ is that without the atoms.  The optical depth in the
plateau is extracted from a fit to the peak in the histogram. We
then fit the variation of peak optical depth assuming
$OD=n\sigma/(1+M_0/M_{sat})$ to determine the depth in the zero
intensity limit $M_0\rightarrow 0$, and thus the surface density of the sample. Here,
$\sigma$=0.347~$\mu$m$^2$ is the known cesium atom-photon
cross-section while the fit parameter $M_{sat}$ represents the
photon number on a CCD pixel at the atomic saturation intensity.

\subsection*{Fluctuation of atomic density}

The fluctuations in the absorption images are estimated by taking
the average of 11 images under the same experimental procedure, and
calculating the mean and variance of optical depth measured at each
CCD pixel. Fluctuations are presumed to arise from optical shot
noise, thermal atomic fluctuation, and long lengthscale variations
arising from total atom number fluctuation. The optical shot noise
is calibrated by examining regions with negligible atomic density,
and extended to higher optical depth using $\delta
OD_{os}\propto\sqrt{1+e^{OD}}$.  For the thermal cloud, with density
$n<0.3$ atoms/site, the fluctuation-dissipation theorem predicts
$\delta N_a = \sqrt{N_a}$, with $N_a$ the number of atoms measured
in a given region. This result should be valid for a region
significantly larger than the correlation length, which we expect
for the normal gas to be on order of the deBroglie thermal
wavelength, expected to be $<1.5\mu m$ for our
sample.  Though each imaging pixel corresponds to an area in the
object plane consisting of ~14 sites, imperfect imaging resolution
is expected to effectively average away a certain fraction of the
total fluctuation. This effect can be calculated, assuming
statistical independence for each site, by summing the weight
$w_{i,j}$ of a resolution-limited spot falling within a given pixel
j for each lattice site i, giving a variance reduced by $\sum_i
w_{i,j}^2$.  The result for our parameters is a reduction to $\delta
n = \gamma\sqrt{n}$, with $\gamma \sim 0.11(1)$.  This should be
compared with the fraction of the total fluctuation shown in Fig.
\ref{fig4} corresponding to thermal fluctuations in the superfluid regime.  To make this
comparison, we reject global fluctuations associated with variation
of the total atom number by subtracting the variance we
calculate after first applying a resolution-spoiling gaussian blur
to the images from the variance without modification.  We find, for the remaining high spatial-frequency
fluctuations, a best fit to $\gamma$ of 0.15(2), using a
gaussian blur $1/e^2$ radius of $r_b=14\mu$m to remove global variations (the result varies within stated error for blur radii $7\mu$m$<r_b<28\mu$m).  The
remaining discrepancy is likely due to calibration of
imaging resolution, and possibly the effect of a nonnegligible
correlation length.

\subsection*{References and Notes}

\begin{enumerate}
\item C. L. Hung, X. B. Zhang, N. Gemelke, C. Chin, {\it Physical Review A}, {\bf 78}, 011604 (2008).
\end{enumerate}

\clearpage

\end{document}